\documentclass[%
 reprint,
 amsmath,amssymb,
 aps,
pra,
]{revtex4-2}

\usepackage{graphicx}
\usepackage{dcolumn}
\usepackage{bm}

\begin{document}
\preprint{APS/123-QED}

\title{Phase retrieval based on shaped incoherent sources}
\author{Ziyan Chen$^{1,2}$}
\author{ Heng Wu$^{1,2}$}
\email{heng.wu@foxmail.com}
\author{ Jing Cheng$^{3}$}
\affiliation{$^{1}$Guangdong Provincial Key Laboratory of Cyber-Physical System, School of Automation, Guangdong University of Technology, Guangzhou 510006, China\\
$^{2}$School of Computer, Guangdong University of Technology, Guangzhou 510006, China\\ 
$^{3}$School of Physics, South China University of Technology, Guangzhou, 510641, China}

%
%
%

\date{\today}

\begin{abstract}
The current ghost imaging phase reconstruction schemes require either complex optical systems,  Fourier transform steps, or iterative algorithms, which may increase the difficulty of system design, cause phase retrieval error or take too much time. To address this problem, we propose a five-step phase-shifting method in which no complex optical systems, Fourier transform steps, or iterative algorithms are needed. With five designed incoherent sources, one can obtain five different corresponding ghost imaging patterns, then the phase information of the object can be calculated from those five speckle patterns. The applicability of this theoretical proposal is demonstrated via numerical simulations with two kinds of complicated objects, and the results illustrate the phase information of the complicated object can be reconstructed successfully and quantitatively.  
\end{abstract}

\maketitle


\section{Introduction}
Ghost imaging technique is a novel imaging method that can non-locally retrieve information of an unknown object with correlation measurements of the intensity fluctuations in two detectors, one detector we call it test detector which has no resolution and usually is a bucket detector or point detector in the test path, the other detector we call it reference detector which has high resolution in the reference path \cite{JCSSH,JC1}.

A lot works based on ghost imaging focus on retrieving the amplitude of an unknown object with its phase neglected. In recent years, some methods were reported to obtain the phase distributions of the unknown object in GI. In 2006, the first phase retrieval method based on GI was reported in which both the amplitude and phase information of the unknown object were extracted from the measurement of the first-order spatial correlation function by use of a modified Young interferometer \cite{RBFG}. In 2008, another GI method based on a large number of numerical iteration calculations has also successfully reconstructed the phase distributions \cite{GRYQW}.  Gong demonstrated a modified GI system in which two mirrors and two beam splitters(BS) were inserted into the standard GI scheme, and find it can separately and non-locally reconstruct both the complex-valued object and its amplitude-dependent part in 2010 \cite{WLGSSH}. Later, A ghost hologram was used to record with a single-pixel configuration by adapting concepts from standard digital holography to extract the intensity and phase information of a structured and realistic object \cite{PCVDET}. Recently, A modified ghost diffraction method combined with the four-step phase-shifting method was reported to extract the relative phase information of a double-slit with a fixed phase difference between the two slit and a Gaussian phase distribution plate \cite{CZY,CZY2}. As for work in \cite{WLGSSH} , it need complicated optical system. And most phase retrieval works  in which the Fourier transform step\cite{RBFG,GRYQW,PCVDET,CZY, CZY2, FT1}  or the iterative algorithm\cite{diedai1,diedai2,diedai3} is very necessary  in the process of reconstructing phase information, which may cause the phase retrieval error or take too much time. So in this paper, we introduce a five-step phase-shifting method which can directly get the phase information of complicated objects without the need of complicated GI system, iterative algorithm and Fourier transform step.

The paper is organized as follows. In section 2, we give the theoretical model. By designing five different incoherent sources, we can retrieve the phase via direct and simple calculations in GI. In section 3, we choose two examples to show the performance of this phase retrieve proposal. Finally in section 4, we give the conclusion.

\section{Model and theory}
\begin{figure}[ht]
\centering
{\includegraphics[width=\linewidth]{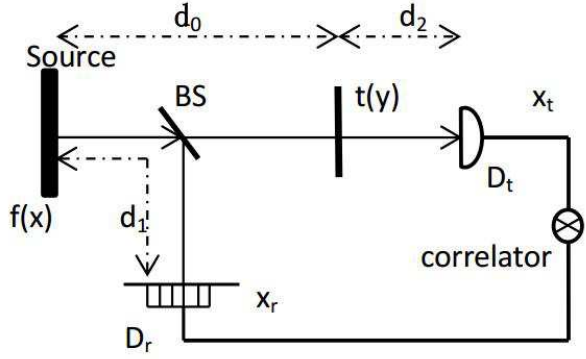}}
\caption{Geometry of the standard GI system. The source field $f(x)$ is split into two beams by the nonpolarized BS.
$x$, $y$, $x_t$, $x_r $ are the positions at the source plane, unknown object plane, test detector plane and reference detector plane.}
\label{f111}
\end{figure}

Our GI scheme is presented in Fig.\ref{f111}, we can see that it is the same as the standard GI system. The classical stochastic optical field is split into two beams by the nonpolarized beam splitter (BS), those two beams propagate through two paths in which we call them reference path and test path. The test path contains an unknown object and a point detector $D_t$, with $d_2$ from the object to the test detector $D_t$ and $d_0$ from the source to the object. The reference path is irrelevant with the object and including a high resolution detector $D_r$, and the distance between the source and the detector $D_r$ is $d_1$.
The intensity distributions recorded by the $D_t$ and $D_r$ are correlated by a correlator to get the GI patterns.

Based on the classical optical coherent theory \cite{JWGITFO}, and the fact that the field fluctuations of a classical light source can be modeled by a complex circular Gaussian random process with zero mean \cite{JWGSO}, so we can get the correlation of the intensity fluctuations between the two detectors 
\begin{eqnarray}
\begin{aligned}
G(x_t,x_r) =&\langle E_r(x_r)E_r^{*}(x_r)E_t(x_t)E_t^{*}(x_t)\rangle \\&-\langle E_t(x_t)E_t^{*}(x_t)\rangle \langle E_r(x_r)E_r^{*}(x_r)\rangle\\
\label{1}
\end{aligned}
\end{eqnarray}
in which $E_t(x_t)(E_r(x_r))$ is the optical field of the test(reference) detector. The field $E_t(x_t)$ can be calculated by the Fresnel integral \cite{JWGITFO}

\begin{eqnarray}
\begin{aligned}
E_t(x_t)=&\frac{-1}{ \lambda \sqrt {d_0 d_2}} \int dy dx_1 E_s(x_1) \exp[-\frac{i\pi}{\lambda d_0}(x_1-y )^2]\\
 &\times t(y)\exp[-\frac{i\pi}{\lambda d_2}(y-x_t)^2]
 \label{2}
\end{aligned}
\end{eqnarray}

Then we can get the field $E_r(x_r)$ similarly 
\begin{eqnarray}
E_r(x_r)=&\frac{1}{\sqrt{i\lambda d_1}} \int dx_2 E_s(x_2) \exp[-\frac{i\pi}{\lambda d_1}(x_2-x_r )^2]
\label{3}
\end{eqnarray}
where $E_s(x_1)(E_s(x_2))$ is the optical field in the source plane.

Combined Eq.\ref{2} and Eq.\ref{3} into Eq.\ref{1}

\begin{eqnarray}
\begin{aligned}
G(x_t,x_r) =&\frac{1}{ \lambda^{3} d_0 d_1 d_2} |\int dy dx_1 dx_2 \Gamma(x_1,x_2) \\
&\times \exp[\frac{i\pi}{\lambda d_1}(x_2-x_r )^2]
\exp[-\frac{i\pi}{\lambda d_0}(x_1-y)^2] \\
&\times t(y)\exp[-\frac{i\pi}{\lambda d_2}(y-x_t)^2]|^2
 \label{4}
\end{aligned}
\end{eqnarray}
in which $\Gamma(x_1,x_2)$ represents the coherence function of the source.

Assuming the source is fully spatial incoherent, then we have
\begin{equation}
\Gamma(x_1,x_2)=\langle E_s(x_1) E_s^{*}(x_2)\rangle =f(x_1)\delta(x_1-x_2)
\label{5}
\end{equation}
where $f(x_1)$ is the intensity distribution of the source.Therefore, Eq.\ref{4} can be rewritten as

\begin{eqnarray}
\begin{aligned}
G(x_t,x_r) =&\frac{1}{ \lambda^{3} d_0 d_1 d_2} |\int dy dx f(x) \exp[\frac{i\pi}{\lambda d_1}(x-x_r )^2]\\
 &\times \exp[-\frac{i\pi}{\lambda d_0}(x-y)^2] t(y)
 \\
 &\times \exp[-\frac{i\pi}{\lambda d_2}(y-x_t)^2]|^2
 \label{6}
\end{aligned}
\end{eqnarray}

To realize GI, the distance are set $d_0=d_1$, and a point detector is placed in the test path, then we can get the GI formula

\begin{eqnarray}
\begin{aligned}
I(x_r)=&G(x_t=0,x_r)\\
=&\frac{1}{ \lambda^{3} d_0^2 d_2}|\int f(x) \exp[\frac{i\pi}{\lambda d_0}2x(y-x_r)]dx\\
&\times \int t(y)\exp[-\frac{i\pi}{\lambda}(\frac{1}{d_0}+\frac{1}{d_2})y^2] dy |^2\\
=&\frac{1}{ \lambda^{3} d_0^2 d_2} \left|
 \int dy F(\frac{x_r-y}{\lambda d_0}) t(y)\exp[-\frac{i\pi}{\lambda}(\frac{1}{d_0}+\frac{1}{d_2})y^2]\right|^2\\
=&\frac{1}{ \lambda^{3} d_0^2 d_2}\left |
 F(\frac{x_r}{\lambda d_0})\otimes  \left(t(x_r)\exp[-\frac{i\pi}{\lambda}(\frac{1}{d_0}+\frac{1}{d_2}){x_r}^2]\right)\right|^2\\ 
 \label{7}
\end{aligned}
\end{eqnarray}
in which F(.) is the Fourier transform of f(.),and $\otimes$ denotes convolution.

For convenience, we use object $tt(x_r)$ to replace $t(x_r)\exp[-\frac{i\pi}{\lambda}(\frac{1}{d_0}+\frac{1}{d_2}){x_r}^2]$, so we have 

\begin{eqnarray}
tt(x_r)=t(x_r)\exp[-\frac{i\pi}{\lambda}(\frac{1}{d_0}+\frac{1}{d_2}){x_r}^2]
\label{8}
\end{eqnarray}
from Eq.\ref{8} we can directly get the conclusion that the amplitude of $tt(x_r)$ is just the same as $t(x_r)$, their phase difference which is related to $x_r$ is also explicit . So if we obtain the phase and amplitude information of the object $tt(x_r)$, then we can reconstruct the phase and amplitude distributions of  $t(x_r)$. 
With Eq.\ref{8} we can rewrite Eq.\ref{7} as
\begin{eqnarray}
I(x_r)=\frac{1}{ \lambda^{3} d_0^2 d_2}|F(\frac{x_r}{\lambda d_0})\otimes tt(x_r)|^2
\label{9}
\end{eqnarray}
we can see that Eq.\ref{9} essentially depicts a coherent imaging system, the object $tt(x_r)$ goes through a coherent imaging system whose point spread function(PSF) is $F(\frac{x_r}{\lambda d_0})$, so we can get the corresponding measurable intensity distributions $I(x_r)$. Mathematically, one can obtain enough incoherent intensity distributions with the help of several different PSF to make the reconstruction of the phase information of the object $tt(x_r)$ become reality. After many attempts, we find five different sources(corresponding to five different PDF) are appropriate and needed to realize our phase retrieval aim.  We design the intensity distributions of the sources as
$f_m(x)=1+u_m e^{(i2 \pi \varepsilon x)}+u_m^{*} e^{(-i2\pi \varepsilon x)}$($m=1,2,3,4,5$, $\varepsilon$ is a real constant, and $u_1=\frac{1}{2},u_2=\frac{1+i}{2 \sqrt{2}},u_3=\frac{i}{2},u_4=\frac{-1+i}{2 \sqrt{2}},u_5=-\frac{1}{2}$ ) to make the $f_m(x)$  a positive real function, $f_m(x)$ is the intensity distributions of the source, so it must be a positive and real. With those five sources, we can get the five corresponding GI patterns

\begin{eqnarray}
I(x_r)=&\frac{1}{ \lambda^{3} d_0^2 d_2}|tt(x_r)+u_m tt(x_r-\varepsilon\lambda d_0)\\ \nonumber
& + u_m^{*} tt(x_r+\varepsilon\lambda d_0)|^2
\label{10}
\end{eqnarray}
After some calculations, we have

\begin{widetext}
\begin{eqnarray}
\begin{aligned}
&\frac{(2 \sqrt{2}-2)(1+i)}{2 \sqrt{2}-4}I_1(x_r)+\frac{[-2 \sqrt{2}+i(2 \sqrt{2}-4)]}{2 \sqrt{2}-4}I_2(x_r)+\frac{4}{2 \sqrt{2}-4}I_3(x_r)+\frac{[-2 \sqrt{2}-i(2 \sqrt{2}-4)]}{2 \sqrt{2}-4}I_4(x_r)\\&+\frac{(2 \sqrt{2}-2)(1-i)}{2 \sqrt{2}-4}I_5(x_r)=tt(x_r+\varepsilon\lambda d_0)tt^{*}(x_r-\varepsilon\lambda d_0)
\label{11}
\end{aligned}
\end{eqnarray}

For connivance, we define

\begin{eqnarray}
\begin{aligned}
H(x_r)=&\frac{(2 \sqrt{2}-2)(1+i)}{2 \sqrt{2}-4}I_1(x_r)+\frac{[-2 \sqrt{2}+i(2 \sqrt{2}-4)]}{2 \sqrt{2}-4}I_2(x_r)+\frac{4}{2 \sqrt{2}-4}I_3(x_r)+\frac{[-2 \sqrt{2}-i(2 \sqrt{2}-4)]}{2 \sqrt{2}-4}I_4(x_r)\\
&+\frac{(2 \sqrt{2}-2)(1-i)}{2 \sqrt{2}-4}I_5(x_r)
\label{12}
\end{aligned}
\end{eqnarray}
\end{widetext}

so we can rewrite Eq.\ref{11} as
\begin{eqnarray}
H(x_r)=tt(x_r+\varepsilon\lambda d_0)tt^{*}(x_r-\varepsilon\lambda d_0)
\label{13}
\end{eqnarray}

It's natural to get
\begin{eqnarray}
\Phi _{H}({x_r}) = \Phi_{tt} (x_r+\varepsilon\lambda d_0)- \Phi_{tt} (x_r-\varepsilon\lambda d_0)
\label{14}
\end{eqnarray}
where $\Phi_{H}(x_r)$ ($\Phi_{tt}(x_r)$) is the phase of $H(x_r)$ ($tt(x_r)$). One can assume the phase at $tt=0$ is zero, i.e., $\Phi_{tt}(0)=0$, then this equation can give the values of $\Phi_{tt}(0)$, $\pm\Phi_{tt}(2\varepsilon\lambda d_0)$, $\pm\Phi_{tt}(\varepsilon\lambda d_0)$,...Now it's obvious that we have reconstructed the phase of $tt(x_r)$ successfully with the phase information of $H(x_r)$. So the next step is to get the phase distributions of $t(x_r)$ (This is exactly what we need to obtain in our scheme) from the phase values of $tt(x_r)$, with Eq.\ref{8} we can directly 
\begin{eqnarray}
\Phi _{t}({x_r}) = \Phi_{tt}(x_r)
+\frac{i\pi}{\lambda}(\frac{1}{d_0}+\frac{1}{d_2}){x_r}^2
\label{15}
\end{eqnarray}
in which $\Phi_{t}(x_r)$ is the phase of $t(x_r)$. So with Eq.\ref{15} and the values of $\Phi_{tt}(0)$, $\pm\Phi_{tt}(2\varepsilon\lambda d_0)$, $\pm\Phi_{tt}(4\varepsilon\lambda d_0)$,... one can directly get the values of $\Phi_{t}(0)$, $\pm\Phi_{t}(2\varepsilon\lambda d_0)$, $\pm\Phi_{t}(4\varepsilon\lambda d_0)$,... Noting that we have directly obtain the phase of the object $t(x_r)$ without the need of Fourier transform step, but in our former work \cite{CZY,CZY2}, we have gotten the phase of the Fourier-transform spectrum of the object, so it's very necessary to use the Fourier transform step to reconstruct the phase information of the unknown object in our former work.  Since we have assumed the phase at $tt(x_r)=0$ is zero, thus say $\phi_{tt}(0)=0$, so the reconstructed phase will have a constant value difference from the real phase, but this absolute phase is not significant for the relative phase distribution keeps invariant.

\section{Numerical simulations}
In the following discussions, we choose two kinds of different complicated objects(four-slit and double-slit Gaussian plate) to verify our GI scheme's validity. In our simulations, we set the source transverse size $D_s$=10mm, and wavelength $\lambda=628 nm$, the distance between the source and the object in the test path is $d_0=400mm$, the distance between the object and the test detector $D_t$ is $d_2=200mm$. The resolution of the CCD is $\Delta_{x_r}=8.3 \mu m$, sample number is M=320, and we set $\varepsilon=3.3041\times 10^{-5} nm^{-1}$.

\begin{figure}[htpb]
\centering
{\includegraphics[width=\linewidth]{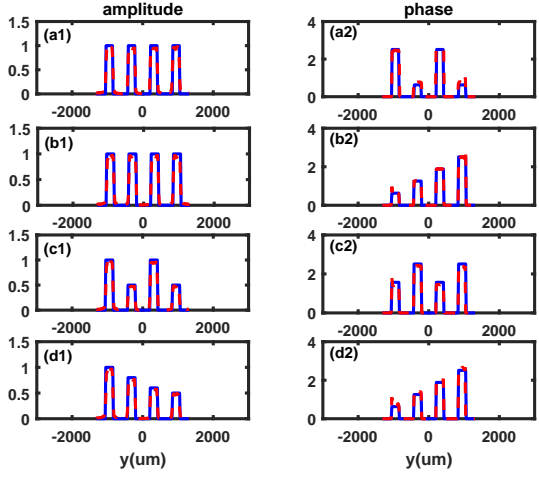}}
\caption{The retrieved phases and amplitudes of the four different four-slits. The blue solid curves represent original phases and amplitudes, the dashed red curves display the reconstructed phases and amplitudes.
(a1)-(a2): The original amplitude and phase of the first four-slit whose parameters are $\alpha_1=\alpha_2=\alpha_3=\alpha_4=1$, $\theta_{\alpha1}=\theta_{\alpha3}=0.8\pi$, $\theta_{\alpha2}=\theta_{\alpha4}=0.2\pi$.
(b1)-(b2): The original amplitude and phase of the second four-slit whose parameters are $\alpha_1=\alpha_2=\alpha_3=\alpha_4=1$, $\theta_{\alpha1}=0.2\pi, \theta_{\alpha2}=0.4\pi$, $\theta_{\alpha3}=0.6\pi, \theta_{\alpha4}=0.8\pi$.
(c1)-(c2): The original amplitude and phase of the third four-slit whose parameters are $\alpha_1=\alpha_3=1, \alpha_2=\alpha_4=0.5$, $\theta_{\alpha1}=\theta_{\alpha3}=0.5\pi$, $\theta_{\alpha2}=\theta_{\alpha4}=0.8\pi$.
(d1)-(d2):  The original amplitude and phase of the fourth four-slit whose parameters are $\alpha_1=1, \alpha_2=0.8, \alpha_3=0.6, \alpha_4=0.5$, $\theta_{\alpha1}=0.2\pi, \theta_{\alpha2}=0.4\pi$, $\theta_{\alpha3}=0.6\pi, \theta_{\alpha4}=0.8\pi$. }
\label{f333}
\end{figure}

The transmittance  of the four-slit is
\begin{equation*}
 t_1(y)=\left\{
 \begin{array}{rcl}
\alpha_1 e^{i\theta_{\alpha1}} & & {-\frac{4w+3d}{2}\leq y \leq  -\frac{2w+3d}{2}}\\
\alpha_2 e^{i\theta_{\alpha2}} & & {-\frac{2w+d}{2}\leq y \leq  -\frac{d}{2}}\\
\alpha_3 e^{i\theta_{\alpha3}} & & {\frac{d}{2}\leq y \leq  \frac{2w+d}{2}}\\
\alpha_4 e^{i\theta_{\alpha4}} & & {\frac{2w+3d}{2}\leq y \leq  \frac{4w+3d}{2}}\\
0  & & {other}\\
\end{array} \right.
\end{equation*}

\begin{figure}[htb]
\centering
{\includegraphics[width=\linewidth]{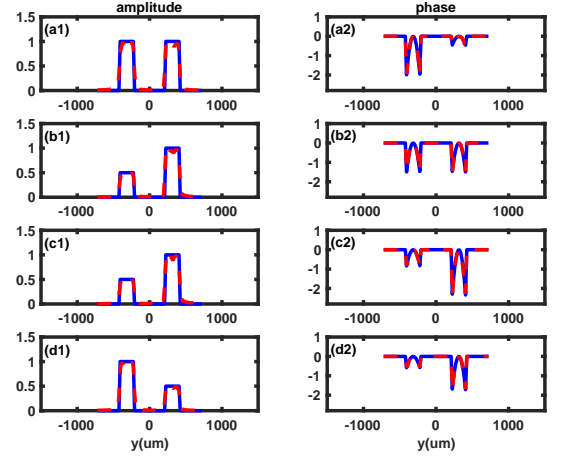}}
\caption{The retrieved phases and amplitudes of the four different double-slit Gaussian plates. The blue solid curves represent original phases and amplitudes, the dashed red curves display the reconstructed phases and amplitudes.
(a1)-(a2): The original amplitude and phase of the first double-slit Gaussian plate whose parameters are $\gamma_1=\gamma_2=1$, $b_1=65 \mu m$, $b_2=135 \mu m$.
(b1)-(b2): The original amplitude and phase of the second double-slit Gaussian plate whose parameters are $\gamma_1=0.5$,$\gamma_2=1$, $b_1=b_2=75 \mu m$.
(c1)-(c2): The original amplitude and phase of the third double-slit Gaussian plate whose parameters are $\gamma_1=0.5$,$\gamma_2=1$, $b_1=100 \mu m$, $b_2=60 \mu m$.
(d1)-(d2):  The original amplitude and phase of the fourth double-slit Gaussian plate whose parameters are $\gamma_1=1$,$\gamma_2=0.5$, $b_1=120 \mu m$, $b_2=70 \mu m$. }
\label{f222}
\end{figure}

in which $w$ represents the slit width, $d$ represents the slit distance, $\alpha_i$ and $\theta_{\alpha i}$ are the amplitude and phase of the $i_{th}$ slit respectively. And we set $w=210$$\mu$m, $d=420$$\mu$m.
Here we choose four different four-slits to demonstrate dependability of our GI method, the reconstructed phases and amplitudes of the four-slits have been displayed in Fig.\ref{f333}. The first four-slit shown in Fig.\ref{f333} (a1-a2) whose parameters are $\alpha_1=\alpha_2=\alpha_3=\alpha_4=1$, $\theta_{\alpha1}=\theta_{\alpha3}=0.8\pi$, $\theta_{\alpha2}=\theta_{\alpha4}=0.2\pi$, it has the same  amplitude in every slit ,  its retrieval phase and amplitude results are reconstructed successfully just as expected.
The second four-slit shown in Fig.\ref{f333} (b1-b2) whose parameters are $\alpha_1=\alpha_2=\alpha_3=\alpha_4=1$, $\theta_{\alpha1}=0.2\pi, \theta_{\alpha2}=0.4\pi$, $\theta_{\alpha3}=0.6\pi, \theta_{\alpha4}=0.8\pi$, its amplitude is just the same as the first four-slit, its phase distribution in every slit is different from each other. The third four-slit shown in Fig.\ref{f333} (c1-c2) whose parameters are $\alpha_1=\alpha_3=1, \alpha_2=\alpha_4=0.5$, $\theta_{\alpha1}=\theta_{\alpha3}=0.5\pi$, $\theta_{\alpha2}=\theta_{\alpha4}=0.8\pi$,   its amplitude and phase also have been successfully reconstructed .  The fourth four-slit shown in Fig.\ref{f333} (d1-d2) is a very complicated for its phase and amplitude are different in every slit, its parameters are $\alpha_1=1, \alpha_2=0.8, \alpha_3=0.6, \alpha_4=0.5$, $\theta_{\alpha1}=0.2\pi, \theta_{\alpha2}=0.4\pi$, $\theta_{\alpha3}=0.6\pi, \theta_{\alpha4}=0.8\pi$, and it is also have been reconstructed successfully under our GI scheme.   The relative difference between the reconstructed phases and the original phases 
is near zero, which  consistent with our theory that have demonstrated in Section 2. In order to get further verification of our GI scheme, we have also simulated  another more complicated kind of object(double-slit Gaussian plate) than the four-slits for the second example.

The second example shown in Fig.\ref{f222} is a double-slit Gaussian plate
\begin{equation*}
 t_2(y)=\left\{
 \begin{array}{rcl}

\gamma_1 e^{-\frac{ i    y^2}{b_1^2}} & & {-\frac{2w+d}{2}\leq y \leq  -\frac{d}{2}}\\
\gamma_2 e^{-\frac{ i   y^2}{b_2^2}} & & {\frac{d}{2}\leq y \leq  \frac{2w+d}{2}}\\

0  & & {other}\\
\end{array} \right.
\end{equation*}

in which $\gamma_i$ is the amplitude of the $i_{th}$  Gaussian plate slit, while $b_i$ represents the width parameter of the phase distribution for the $i_{th}$   Gaussian plate slit. It's also very clear that both  amplitudes and phases of the  double-slit Gaussian plate have been reconstructed successfully in Fig.\ref{f222}, and the relative difference between the reconstructed phases and the original phases are also near zero, which once again demonstrates the validity of our GI phase retrieval scheme.

\section{Conclusion}
In conclusion,we provide an algorithm which we called it five-step phase-shifting method to reconstruct the phase distribution of the unknown object based on ghost imaging, this method does not need any complicated GI system, iterative algorithms or Fourier transform step. To demonstrate the validity of our GI scheme, we have simulated two kinds of complicated objects(four-slit and double-slit Gaussian plate), the results demonstrate that our scheme  is  very dependable.

\begin{acknowledgments}

The authors thank the supports from the National Natural Science Foundation of China(Grant No.62173098, 11774097, U20A6003, U2001201, U1801263, 61805048),Key-Area Research and Development Program of Guangdong Province (Grant No.2021B0101220001, 2019B030330001),Guangdong Provincial Key Laboratory of Cyber-Physical System (2020B1212060069).

\end{acknowledgments}

\end{document}